\documentclass [12 pt]{article}
\def\be{\begin{equation}}
\def\eea{\end{eqnarray}}
\def\bea{\begin{eqnarray}}
\def\ee{\end{equation}}

\def\rm{\bf}
\def\iota{\imath}
\usepackage[psamsfonts]{amssymb}
\usepackage{amsmath}
\author{M. Amooshahi$^{1}$
\footnote{amooshahi@sci.ui.ac.ir} ,  E. Amooghorban$^{1}$
\footnote{amooghorban@sci.ui.ac.ir}\\
{\small $^{1}$Department of Physics, Faculty of Science, University
of Isfahan,}\\{\small Hezar Jarib Ave., Isfahan, Iran}}

\title{Canonical quantization of a  dissipative system interacting with an   anisotropic non-linear absorbing environment }
\begin{document}
\maketitle
\begin{abstract}
A canonical quantization scheme is represented for  a  quantum
system interacting with a nonlinear absorbing environment. The
environment is taken  anisotropic and the main system is coupled  to
its environment through some coupling tensors of various ranks. The
nonlinear response equation of the environment against  the motion
of the main system is obtained.  The nonlinear
Langevin-schr\"{o}dinger  equation is concluded as the macroscopic
equation of motion of the dissipative system. The effect of
nonlinearity of the environment is investigated on the spontaneous
emission of an initially excited two level-atom imbedded in such an environmrnt.\\
\\
{\bf Keywords:  Dissipative  system, Absorbing environment,
nonlinear response equation, Coupling tensor, Susceptibility tensor,
Canonical quantization, Nonlinear Langevin-schr\"{o}dinger equation, Spontaneous emission.}\\
\\
{\bf PACS number: 42.50.Ds, 03.65.-w, 03.70.+k}
\end{abstract}
\section{Introduction}

The simplest way of describing a damped system in classical dynamics
is by adding a resisting force, generally velocity-dependent, to the
equation of motion of the system. Frequently the magnitude of the
resisting force may be closely presented, over a limited range of
velocity, by the law $f_d=av^n$, where $v$ is the velocity of the
damped system and $a$ and $n$ are constants. For example for the
friction force $n=0$, viscous force $n=1$ and for high speed motion
$n=2$ \cite{1}. Such an approach is no longer possible in quantum
mechanics, because one can not find a unitary time evolution
operator for both the states and
the observables,  consistently.\\
In order to take into account the dissipation in a quantum system,
there are usually two approaches. The first approach is a
phenomenological way, by which the effect of dissipation is taken
into account by constructing  a suitable Lagrangian or Hamiltonian
for the system \cite{2,3}. Following this method the first
Hamiltonian was proposed by Caldirola \cite{4} and   Kanai \cite{5}
and afterward by others \cite{6,7}. There are  difficulties about
the quantum mechanical solutions of the Caldirola-Kanai Hamiltonian.
For example quantization using this way violates the uncertainty
relations or canonical commutation rules. The uncertainty  relations
vanishes as time tends to infinity \cite{8}-\cite{11}. \\

The second approach is based on the assumption that the damping
forces  is caused by an irreversible transfer of  energy from the
system to a reservoir \cite{12,13}. In this method ,  modeling the
absorptive environment by a collection  of harmonic oscillators and
choosing a suitable interaction between the system and the
oscillators, a consistent quantization is achieved for both the main
system and the environment\cite{14}-\cite{26}. In the Heisenberg
picture, one can obtain  the  linear Langevin-schr\"{o}dinger
equation, as the macroscopic equation of motion of the main
system.\cite{14,15}.

In the present work, following the second approach,  a fully
canonical quantization is introduced for a system moving in an
  anisotropic non-linear absorbing environment. The dissipative
system is the prototype of some important problems which the present
approach can be applied to cover such problems straightforwardly.

The  paper is organized as follows: In section 2, a Lagrangian for
the total system (the main system and the environment) is proposed
and a classical treatment of the  dissipative system is achieved. In
section 3,  the Lagrangian introduced in the section 2 is used for a
canonical quantization of both the main system and the non-linear
environment. In section 4,  the present quantization is used to
investigate  the effect of the nonlinearity of the  environment on
the spontaneous decay rate of an initially excited two-level atom
embedded in the absorbing environment.  Finally, the paper is closed
with a summary and some concluding remarks in section 5.
\section{Three-dimensional quantum dissipative systems}
 When an absorbing environment responds  non-linearly against  the motion of a  system, the
non-linear Langevin-Schr\"{o}dinger equation is usually appeared as
the macroscopic equation of motion of the system. As an example,
when the electromagnetic field  is propagated in  an absorbing
non-linear polarizable medium,   the vector potential satisfies the
non-linear Langevin-Schr\"{o}dinger equation. In this section, the
motion of a three-dimensional system in the presence of an
anisotropic non-linear  absorbing environment is classically
treated. For this purpose , the environment is modeled by a
continium of three dimensional harmonic oscillators labeled by a
continuous parameter $\omega$.  The total Lagrangian is proposed as
\begin{equation} \label{a1}
L(t) = L_{e}  + L_{s} +L_{int}.
\end{equation}
which is the sum of three pars. The part $L_{e}$ is the Lagrangian
of the environment
\begin{equation} \label{a2}
L_{e}(t) \, = \int_0^\infty  d\omega \left[\frac{1}{2}{\rm \dot
X}(\omega ,t)\cdot {\rm \dot X}(\omega,t)   - \frac{1}{2}\omega
^2\ {\rm X}(\omega,t) \cdot {\rm X}(\omega,t) \right].
\end{equation}
where ${\rm X}(\omega,t)$ is the dynamical variable of the
oscillator labeled by $\omega $.  The second part $L_{s}$ in
(\ref{a1}) is the Lagrangian of the main system. Taking the system
as a particle with mass, $m$, moving under an external potential
$V(\rm q)$, one  can write
\begin{equation} \label{a3}
L_{s} \, = \frac{1}{2}m\ {\rm \dot q}(t)   \cdot {\rm \dot q}(t) -
V(\rm q).
\end{equation}
The last part  $ L_{int} $ in the total Lagrangian (\ref{a1}) is
 the interaction term between the system and its absorbing
environment and includes both the linear and nonlinear
contributions as follows
\begin{eqnarray} \label{a4}
&& L_{int}  = \int_0^\infty  d\omega \,f_{ij}^{(1)} (\omega )
\dot q^i (t)\,X^j(\omega,t)  \nonumber\\
&&+\int_0^\infty {d\omega } \int_0^\infty  {d\omega'}
f_{ijk}^{(2)} (\omega ,\omega
')\ \dot {q}^i (t)\,{X}^j(\omega,t) {X}^k(\omega' ,t)\nonumber\\
&&+\int_0^\infty  {d\omega }\int_0^\infty  {d\omega
'}\int_0^\infty {d\omega ''}\,f_{ijkl}^{(3)}
(\omega,\omega',\omega'' )\dot {q}^i (t)\,{ X}^j(\omega
,t){X}^k(\omega' ,t) { X}^l(\omega'',t)+\dots\cdots\nonumber\\
&&
\end{eqnarray}
where $f^{(1)} , f^{(2)}, f^{(3)}, \cdots$  are the coupling tensors
of the main system and its environment. As it is seen from
(\ref{a4}) the coupling tensor $f^{(1)}$ describes the linear
contribution of the interaction part and the sequence $ f^{(2)},
f^{(3)},....$  describe, respectively,  the first order of the
non-linear  interaction part, the second order of the non-linear
interaction part and so on. The interaction Lagrangian (\ref{a4}) is
the generalization of the Lagrangian that previously has been
applied to quantize the electromagnetic field in the presence of
anisotropic
 linear magnetodielectric media \cite{27}.\\
 The coupling tensors  $f^{(1)}
, f^{(2)}, f^{(3)},...$ in  (\ref{a4}) are the key parameters of
this quantization scheme. As it will be seen, in the next section,
the susceptibility tensors of the environment (of the various ranks)
are expressed in terms of the coupling tensors. Also the noise
forces are obtained in terms of the coupling  tensors and the
dynamical variables of the environment at $t=-\infty$.
 \subsection{The classical  Lagrangian equations }
the classical equations of  motion of the total system can be
obtained  using  the principle of the Hamilton's least action, $
\displaystyle \delta \int dt \ \underline{L}(t)=0$. These equations
are the Euler-Lagrange equations. For the dynamical variables, ${\rm
X}(\omega,t)$,  the Euler-Lagrange equations are as
\begin{eqnarray}\label{a5}
&&\frac{d}{{dt}}\left( \frac{\delta  L }{\delta (  \dot {X}_i
(\omega,t))} \right) - \frac{{\delta  L }}{{\delta ( { X}_i
(\omega,t)
 )}} = 0 \hspace{+2cm}i=1,2,3 \nonumber\\
&&\nonumber\\
 && \Rightarrow   \ddot { X}
_i(\omega,t)  + \omega ^2 X _i(\omega,t)  = \dot {q}^j (t) f
_{ji}^{(1)} (\omega )  \nonumber\\
&&\nonumber\\
 &&+\int_0^\infty {d\omega '} \,\dot {q}^j(t)\left[ f
_{jik}^{(2)}(\omega ,\omega ')+ f
_{jki}^{(2)}(\omega',\omega )\right]  X^k (\omega ',t) \nonumber\\
&&\nonumber\\
&& + \int_0^\infty  {d\omega '} \int_0^\infty {d\omega
''}\,\,\dot {q}^j(t) \left[ f _{jikl}^{(3)}(\omega ,\omega
',\omega'')+ f _{jkil}^{(3)}(\omega' ,\omega
,\omega'')\right.\nonumber\\
&&+\left.f_{jkli}^{(3)}(\omega' ,\omega '',\omega)\right] X
^k(\omega',t) X ^l(\omega'',t) +  \ \cdots \cdots
\end{eqnarray}
Also the Lagrange equations for the freedom degrees of the main
system are obtained as follows
\begin{eqnarray}\label{a6}
&& \frac{d}{{dt}}\left( {\frac{{\delta  L }}{{\delta (\dot{q}_i
(t))}}} \right) - \frac{{\delta  L }}{{\delta
(q_i (t))}} = 0\hspace{+2cm}i=1,2,3  \nonumber\\
&&\nonumber\\
 &&\Rightarrow m{\rm \ddot {\rm q}}(t) + \triangledown
V({\rm q}) = - \dot {\rm {R}}(t)
\end{eqnarray}
where
\begin{eqnarray}\label{a7}
 &&R_i(t)= \int_0^\infty  {d\omega } \,  f ^{(1)}_{ij}
(\omega ) \, X^j(\omega ,t)+ \int_0^\infty  {d\omega }
\int_0^\infty  {d\omega '}  f_{ijk} ^{(2)}\, (\omega ,\omega ')
 X^j(\omega ,t) \  X ^k(\omega ',t)\nonumber\\
 &&\nonumber\\
 &&+ \int_0^\infty  {d\omega } \int_0^\infty  {d\omega '}  f_{ijkl}
^{(3)} (\omega ,\omega ',\omega'')\,  X^j(\omega ,t)\
X^k(\omega,t)\  X^l(\omega '',t)+\cdots\nonumber\\
&&
\end{eqnarray}
In Eq. (\ref{a6}) $- \dot {\rm {R}}(t)$ is the force exerted  on the
main system due to its motion  inside the absorbing environment. It
will be seen that the force $- \dot {\rm {R}}(t)$ can be separated
into two parts. One part  is the damping force which is dependent on
the various powers of the velocity of the main system. The second
part is the noise forces which has sinusodial time dependence. Both
the damping  and  the noise forces are necessary for a consistent
quantization of a dissipative system. Without the noise forces the
quantization of a dissipative system encounter inconsistency.
According to the fluctuation- dissipation theorem the absence of any
of these two parts leads to the vanishing of the other part.
\section{Canonical quantization}
In order to represent   a canonical quantization, the canonical
conjugate momenta corresponding to the dynamical variables ${\rm
X}(\omega,t) $ and ${\rm q}$ should be computed  using the
Lagrangian (\ref{a1}). These momenta are as follows
\begin{equation} \label{b1}
 Q _i(\omega ,t)   = \frac{{\delta  L }}{{\delta ( {\dot { X}}
_i(\omega ,t) )}} =  {\dot { X}}_i(\omega,t)  \hspace{1cm} i=1,2,3
\end{equation}
\begin{equation}\label{b1.1}
  {p} _i (t) = \frac{{\delta L}}{{\delta (\dot {q}_i )}} = m\dot {
q}_i + { R}_i (t)\hspace{1cm} i=1,2,3.
\end{equation}
Having the canonical  momenta, both the dissipative system  and the
environment can be quantized in a standard fashion by imposing the
following equal-time commutation rules
\begin{eqnarray} \label{b2}
 &&\left[ { { q}_i (t)\ ,\ { p}_j (t)} \right] = \iota \hbar \delta _{ij}
\end{eqnarray}
\begin{equation}\label{b3}
\left[ { { X} _i(\omega, t)\  ,\ { Q} _j(\omega ',t)} \right] =
\iota \hbar \delta _{ij} \delta (\omega  - \omega ')
\end{equation}
 Using the Lagrangian
(\ref{a1}) and the expressions for the canonical  momenta given by
(\ref{b1}) and (\ref{b1.1}), the Hamiltonian of the total system
clearly can be  written as
\begin{eqnarray} \label{b4}
&&H(t) = \frac{{\left[ {{\rm p}(t) - {\rm R}(t) } \right]}^2}{2m}+
V({\bf q}) +\frac{1}{2} \int_0^\infty  {d\omega } {\left[ {{\rm Q}
^2(\omega,t)  + \omega ^2  {{\rm X}^2 (\omega ,t) }  } \right]}
\end{eqnarray}
where the cartesian components of ${\rm R}(t)$ is defined  by
(\ref{a7}). The Hamiltonian (\ref{b4}) is the counterpart of the
Hamiltonian of the quantized electromagnetic field in the presence
of magnetodielectric media\cite{27}-\cite{29}. Using  the
commutation relations (\ref{b2}), (\ref{b3}) and applying the total
Hamiltonian (\ref{b4}), it can be shown that the combination of the
Heisenberg equations of motion of the canonical variables $ {\rm
X}(\omega,t) $ and $ {\rm Q}(\omega,t)$  leads to the Eq.(\ref{a5}).
Similarly, one can obtain Eq.(\ref{a6})  as the
equation of motion of  ${\rm q}(t)$  in the Heisenberg picture.\\
Let us introduce the  annihilation and creation operators of the
environment as follows
\begin{eqnarray} \label{b4.1}
&& b_i  (\omega ,t) = \sqrt {\frac{1}{{2\hbar \omega }}} \left[
{\omega { X} _i(\omega ,t) + \iota  { Q} _i(\omega , t)} \right].
\end{eqnarray}
From the  commutation relations (\ref{b3}) it is clear that the
ladder operators $ b_i  (\omega ,t)$ and  $b^\dag_i (\omega ,t)$
obey the commutation relations
\begin{eqnarray} \label{b4.2}
&& \left[ {b_i  (\omega ,t),b_{j}^ \dag (\omega ',t)} \right] =
\delta _{ij} \delta (\omega  - \omega ')
\end{eqnarray}
 The Hamiltonian (\ref{b4})  can be rewritten in terms of the creation and annihilation operators  $ b_i  (\omega
 ,t)$
  and $ b^\dag_i  (\omega ,t)$ as follows
\begin{eqnarray} \label{b4.3}
 H = \frac{{\left( {{\rm p} -
{\rm R}(t) } \right)^2 }}{2m} + V({\rm q})  + H_m
\end{eqnarray}
where
\begin{eqnarray} \label{b4.4}
&& H_m  = \sum_{i=1}^3{\int {d\omega } } \,\hbar \omega \,\,b_i ^
\dag (\omega ,t)b_i  (\omega ,t)
\end{eqnarray}
is the Hamiltonian of the  absorbing environment  in the normal
ordering form and
\begin{eqnarray}\label{b4.5}
&&R_i(t) =  {\int_0^\infty {d\omega } } \,\sqrt {\frac{\hbar
}{{2\omega }}}  f _{ij}^{{(1)} } (\omega )\left[ {b_j (\omega ,t) +
 b_j ^ \dag  ( \omega ,t)} \right]  \nonumber\\
 && + { {\int_0^\infty
{d\omega } } \int_0^\infty  {d\omega '} } \frac{\hbar }{{2\sqrt
{\omega \omega '} }}\, f _{ijk}^{{(2)} } (\omega ,\omega ')
 \left[ {b_j  (\omega ,t)b_{k} (\omega ',t) + } +b_j
^ \dag ( \omega ,t)b_{k}^ \dag  (\omega ',t)\right.\nonumber\\
&&\left. +{b_j (\omega ,t)b_{k}^ \dag  ( \omega ',t) + b_j ^ \dag (
\omega ,t)b_{k} (\omega ',t)} \right]+ \cdots
\end{eqnarray}
are the cartesian components of the operator ${\rm R}$, where the
summation should be done over the repeated indices.
\subsection{The response equation of the environment}
 The response equation of the absorbing environment is the base of separating   the force $-{\rm \dot{R}}(t)$ , in the right hand of (\ref{a6}),
 into two parts, that is, the damping force and the noise force. If
 Eq.(\ref{a5})  is solved for ${\rm X}(\omega , t)$ and then,  the obtained  solution is
 substituted  into the definition of ${\rm R}(t)$ given
 by (\ref{a7}), one can obtain the response equation of the
 environment. The  differential equations (\ref{a5}) are a  continuous collection of coupled non-linear
 differential equations for the dynamical variables ${\rm X}(\omega , t)$. The exact solution of
 this equation is impossible unless an iteration method to be used.   For simplicity here we apply
the first order of approximation and  neglect the terms containing
the coupling tensors $f^{(2)} , f^{(3)},...$ in the right hand of
(\ref{a5}) and write  the solution of Eq.(\ref{a5}), approximately,
as
\begin{eqnarray} \label{b8}
{\rm X} (\omega,t)   =  {{\rm  X}}_N (\omega ,t)+ \int_{-\infty}^t
{dt'} \frac{{\sin \omega (t - t')}}{\omega
 }\ (f  ^{(1)})^{\dag} (\omega ) \cdot {\rm \dot
 q}(t^\prime),\nonumber\\
 &&
\end{eqnarray}
where $ (f^{(1)})^{\dag}_{ij} (\omega )= f_{ji} ^{(1)}(\omega)$ and
${{\rm  X}}_N (\omega ,t)$ is the solution of homogeneous equation
$\ddot {\rm X} _N(\omega,t)  + \omega ^2 {\rm X} _N(\omega,t)  =0$.
In fact ${{\rm  X}}_N (\omega ,t)$ is asymptotic form of ${{\rm X}}
(\omega ,t)$ for very large negative times and can be written as
\begin{equation} \label{b8.1}
{{ X}}_{Ni} (\omega ,t)=  \sqrt {\frac{\hbar}{{2 \omega
}}}\left[b^{in}_i(\omega)e^{-\imath\omega
t}+b^{\dag^{in}}_i(\omega)e^{\imath\omega t}\right]
\end{equation}
where $b^{in}(\omega)$ and $b^{\dag^{in}}(\omega)$ are some time
independent annihilation and creation operators which obviously
satisfy the same commutation relations (\ref{b4.2}). The
approximated solution (\ref{b8}) yields   the response equation of
the environment, such that,  the susceptibility tensors appearing in
it, satisfy the
various symmetry properties reported  by the literature \cite{30}.\\
 \\
Now substituting ${\rm X} (\omega,t) $ from (\ref{b8}) in
(\ref{a7}),
 the response equation of the non-linear absorbing environment  is found as
follows
\begin{eqnarray}\label{b9}
&&{\rm R}(t)={\rm R}^{(1)}+{\rm R}^{(2)}+....\nonumber\\
&&{ { R}^{(1)}_i}(t) = \int_{-\infty}^{+\infty}  {dt\,} \chi_{ij}^{(1)} (t - t')  { \dot { q}}_j(t')\, + { R}_{N\,i}^{(1)} (t)\nonumber \\
&&{ { R}^{(2)}_i}(t) =\int_{-\infty}^{+\infty}  {dt'}
\int_{-\infty}^{+\infty} {dt''}\chi_{ijk}^{(2)} (t - t',t - t'')  {
\dot { q}}_j(t')\,{ \dot { q}}_k(t'') + {R}_{N\,i}^{(2)}
(t)\nonumber\\
&&
\end{eqnarray}
where $\chi ^{(1)}$ is the susceptibility tensor of the environment
in the linear regime and is defined by
\begin{eqnarray} \label{b10}
&&\chi ^{(1)}_{ij} (t) =\left\{ \begin{array}{cc}
\displaystyle\int_0^\infty  {d\omega } \frac{{\sin\omega t}}{\omega_1 } \,\, f ^{(1)}_{in} (\omega) f ^{(1)}_{jn} (\omega ) & \hspace{2.00 cm} t>0 \\
\\
 0 & \hspace{2.50 cm} t\leq 0\end{array}\right.
 \end{eqnarray}
 and $ \chi ^{(2)}_{ijk}$  causes  the first order of the
nonlinearity of the response equation,  where for for $t_1,t_2\geq0$
are given by
 \begin{eqnarray}\label{b10.1}
&&\chi ^{(2)}_{ijk} (t_1,t_2) = \int_0^\infty {d\omega_1 }
\int_0^\infty {d\omega_2}\  \frac{\sin \omega_1 t_1}{\omega_1}\
\frac{ \sin \omega_2 t_2}{ \omega_2 } \  f ^{(2)}_{inm} (\omega_1
,\omega_2)\ f ^{(1)}_{jn} (\omega_1 )\ f ^{(1)}_{km} (\omega_2 )\nonumber\\
&&
\end{eqnarray}
 and $\chi ^{(2)}_{ijk}$ is zero for $t_1,t_2<0$.  In (\ref{b10}) and (\ref{b10.1}) the
summation should be done over the repeated indices $m,n$.
 From the definition
(\ref{b10}) it is clear that  $\chi ^{(1)}$ is a symmetric tensor,
$\chi ^{(1)}_{ij} =\chi ^{(1)}_{ji}  $. There are also some symmetry
features  for the non-linear susceptibility tensors of the various
orders. These symmetry properties can be satisfied by imposing some
conditions on the coupling tensors $ f^{(2)} , f^{(3)} , ... $ . For
example the  susceptibility tensor $ \chi ^{(2)}$,
  should satisfy the symmetry property \cite{30}
\begin{equation}\label{b11}
\chi^{(2)}_{ijk}(t_1, t_2)=\chi^{(2)}_{ikj}(t_2 , t_1)
\end{equation}
where is fulfilled   provided that  the coupling tensor $f^{(2)} $
obey the symmetry  condition
\begin{equation}\label{b12}
f^{(2)}_{ijk}(\omega,\omega')= f^{(2)}_{ikj}(\omega',\omega),
\end{equation}
Similarly inserting   the approximated solution ${\bf X}(\omega ,t)$
from (\ref{b8}) into (\ref{a7}), one can obtain the $(n-1)$'th
susceptibility tensor of the environment in the non-linear regime
for $t_1,t_2,...t_n\geq0$ as the following
\begin{eqnarray}\label{b12.1}
&&\chi^{(n)}_{i\ i_1...i_{n}}(t_1,t_2,...,t_{n})=\int_0^\infty
d\omega_1\int_0^\infty d\omega_2....\int_0^\infty d\omega_{n}\
\frac{\sin\omega_1 t_1}{\omega_1}\ \frac{\sin\omega_2
t_2}{\omega_2}....\frac{\sin\omega_{n} t_{n}}{\omega_{n}}\nonumber\\
&&\times\ f^{(n)}_{ij_1j_2....j_{n}}( \omega_1,\omega_2,...\
\omega_n)\ f^{(1)}_{i_1j_1}(\omega_1)\
f^{(1)}_{i_2j_2}(\omega_2)....\ f^{(1)}_{i_nj_n}(\omega_n)
\end{eqnarray}
 and $\chi^{(n)}_{ii_1...i_{n}}(t_1,t_2,...,t_{n})$ is identically  zero for $t_1,t_2,...t_n<0$.
 the Susceptibility tensor $\chi^{(n)}_{ii_1...i_{n}}(t_1,t_2,...,t_{n})$ should  satisfy the symmetry
 relations \cite{30}
\begin{eqnarray}\label{b12.2}
&&\chi^{(n)}_{i\
i_1...i_k,...i_l,...,i_n}(t_1,t_2,...,t_k,...t_l,...,t_{n})
=\chi^{(n)}_{i\ i_1...i_l,...i_k,...,i_n}(t_1,t_2,...,t_l,...t_k,...,t_{n})\nonumber\\
&&
\end{eqnarray}
where this symmetry relation is clearly  fulfilled by imposing the
symmetry conditions
\begin{eqnarray}\label{b12.3}
&&f^{(n)}_{ij_1j_2.......,j_k,...,j_l,...,j_n}(
\omega_1,\omega_2,...,\omega_k,...,\omega_l,...,\
\omega_n)\nonumber\\
=&&f^{(n)}_{ij_1j_2......., j_l, ...,j_k,...,j_n}(
\omega_1,\omega_2,...,\omega_l,...,\omega_k,...,\ \omega_n)
\end{eqnarray}
on the $n$'th coupling tensor in the interaction Lagrangian
(\ref{a4}).\\

 In Eq. (\ref{b9}) ${\rm R}_{N}^{(1)} (t)$ and ${\rm
R}_{N}^{(2)} (t)$ are the noise forces in the linear regime and the
first order of non-linearity , respectively, and  using  the
symmetry relation (\ref{b12}) are obtained as
\begin{eqnarray}\label{b13}
&&R_{N\,i}^{(1)} (t)= \int_0^\infty  {d\omega } \,f_{ij}^{(1)}
(\omega ) { X}_N^j(\omega , t)\nonumber\\
&& {\rm R}_{N\,i}^{(2)} (t) = \int_0^\infty  {d\omega }
\int_0^\infty d\omega ' \,f_{ijk}^{(2)} (\omega ,\omega ')\
X_N^j(\omega , t)\  X_N^k(\omega' , t) \nonumber\\
&&+\int_0^\infty d\omega\ \int_0^\infty d\omega'\ f_{inm}^{(2)}
(\omega ,\omega ')\ f^{(1)}_{jm}(\omega')\ \int_{-\infty}^t \ dt'\
\frac{\sin\omega'(t-t')}{\omega'} \nonumber\\
&&\times \left[ X_N^n( \omega ,t)\ \dot{q}^j(t')+\dot{q}^j(t')\
X_N^m( \omega' ,t)\right]
\end{eqnarray}
where the summation should be done over the repeated indices and
${\bf X}_N(\omega , t) $ is the asymptotic solution (\ref{b8.1}).\\
It is remarkable that for some known susceptibility tensors  $ \chi
^{(1)} ,\chi ^{(2)},...,\chi ^{(n)}$,  the coupling tensors  $ f
^{(1)}, f ^{(2)},... f ^{(n)}$  satisfying  the definitions
(\ref{b10}),(\ref{b10.1}) and (\ref{b12.1}) are not
 unique. In fact if the coupling tensors $ f
^{(1)}, f ^{(2)},..., f^{(n)}$ satisfy  (\ref{b10}),(\ref{b10.1})
and (\ref{b12.1}) for the given susceptibility tensors, also the
coupling tensors $ f'^{(1)}, f'^{(2)},..., f'^{(n)}$ defined by
\begin{eqnarray}\label{b14}
&&f'^{(1)}_{ij}=f ^{(1)}_{im} A_{jm}\nonumber\\
&&f'^{(n)}_{i\ i_1i_2....i_n}=f^{(n)}_{i\ j_1j_2...j_n}\
A_{i_1j_1}A_{i_2j_2}....A_{i_nj_n}
\end{eqnarray}
 satisfy (\ref{b10}),(\ref{b10.1}) and
(\ref{b12.1}), where $ A$ is an orthogonal matrix
$A_{im}A_{mj}=\delta_{ij}$.  The various choices of the coupling
tensors $ f ^{(1)}, f ^{(2)},..., f^{(n)}$ which is related to each
other by the orthogonal transformation (\ref{b14}) do not change the
physical observables. The commutation relations between the
dynamical variables of the total system remain unchanged  under the
orthogonal transformation (\ref{b14}). For example in the next
section it is shown that the decay rate of an initially excited
two-level atom, embedded in a non-linear absorbing environment, are
independent of the various choices of the coupling tensors  which is
related to each other by the transformation (\ref{b14}).\\
 Now  combination of  the response equation (\ref{b9})
and equation (\ref{a6}) yields the non-linear
Lagevin-schr\"{o}dinger equation
\begin{eqnarray} \label{b16}
&& m \ddot{q}_i(t) + \int_{-\infty}^{+\infty} dt'
\dot{\chi}^{(1)}_{ij} (t - t')\ \dot {q}^j(t')+\nonumber\\
&&+\int_{-\infty}^{+\infty} {dt'} \ \int_{-\infty}^{+\infty} {dt''}\
\ddot{\chi}_{ijk} ^{(2)} (t - t',t - t'') \dot {q}^j(t')\ \dot
{q}^k(t'') +....\nonumber\\
&&+\frac{\partial V(\vec{q})}{\partial q^i} = - \dot{R}_{N\ i}^{(1)}
(t)- \dot{R}_{N\ i}^{(2)} (t)+...
\end{eqnarray}
as the macroscopic equation of motion of the main system in the
anisotropic non-linear absorbing environment. The velocity dependent
terms in the left hand of this equation are
 the damping forces exerted on the main system. The noise forces $- \dot{R}_{N\
 i}^{(1)},
 -\dot{R}_{N\ i}^{(2)},...$ in the right hand of (\ref{b16}) are
 necessary for a consistent quantization of the dissipative system.
  As a realization,  if this
quantization method would be applied for the electromagnetic field
in the presence of an absorbing non-linear dielectric medium, the
vector potential would satisfy the equation (\ref{b16}). In that
case, the  tensors $\chi ^{(1)}, \chi ^{(2)},....$ would play the
role of the electric susceptibility tensors and $-
\dot{\rm{R}}_{N}^{(1)}, - \dot{\rm{R}}_{N}^{(2)},...$ would be the
noise polarization densities of various orders.
\section{The effect of nonlinearity of the environment on the spontaneous emission of a two-level atom imbedded in
an  absorbing environment}

In this section the effect of non-linearity of the absorbing
environment is investigated on the spontaneous emission of a
two-level atom embedded in such an environment.  To calculate the
spontaneous decay rate of an initially excited two-level atom, the
quantization scheme in the preceding section is used and the theory
of damping based on the density operator method is applied
\cite{31}. Neglecting the second power of the operator ${\bf R}$ in
(\ref{b4.3}) the Hamiltonian of the total system can be written as
\begin{eqnarray} \label{c1}
&&H = H_0  + H' \nonumber\\
&&H_0  = H_{S}  + H_m=\frac{{\bf p^2}}{2m}+V({\bf
q})+\sum_{i=1}^3\int d\omega   \,\hbar \omega \,\,b_i ^
\dag (\omega)b_i  (\omega) \nonumber \\
&&H' =   -{\rm p} \cdot {\rm  R}
\end{eqnarray}
Let us suppose the main system is  a one electron atom with two
eigenstates $ |1\rangle $ and $ |2\rangle $ correspond to two
eigenvalues $E_1$ and $E_2$, respectively $(E_2>E_1)$.  the
Hamiltonian (\ref{c1}) can now be rewritten as \cite{31}, \cite{32}.
\begin{eqnarray} \label{c2}
&&H = H_0  + H' \nonumber\\
&&H_0  = \hbar\omega_0 \sigma^\dag\sigma +\sum_{i=1}^3\int d\omega
\,\hbar \omega \,\,b_i ^ \dag (\omega)b_i  (\omega)\hspace{2.00 cm}\omega_0=\frac{E_2-E_1}{\hbar}\nonumber\\
&&H'=\imath\ m\omega_0{\bf R}\cdot\left[ {\bf d} \sigma-{\bf
d}^* \sigma^\dag\right]\nonumber\\
&&
\end{eqnarray}
 where $m$ is the electron mass of the atom,  $\sigma=|1 \rangle \langle 2|\ , \sigma^\dag=|2
\rangle\langle1 | $ , are  the Pauli operators  and ${\bf d}=\langle
1 |{\bf r}|2\rangle$, where $ {\bf r}$ is the position vector of the
electron with respect to the center of mass of the atom. Dropping
the energy noncoserving terms correspond to rotating wave
approximation and regarding the relation (\ref{b4.5}), the
interaction term $H'$ up to the first order of nonlinearity  in the
interaction picture is expressed as
\begin{eqnarray} \label{c3}
&&H'_I(t)= e^{\frac{\imath H_0 t}{\hbar}}\ H'(0)\  e^{\frac{-\imath
H_0 t}{\hbar}}\nonumber\\
&&=\imath m\omega_0\int_0^\infty d\omega\
\sqrt{\frac{\hbar}{2\omega}}\ f^{(1)}_{ij}(\omega)\left[d_i\sigma\
b^\dag_j(\omega)\ e^{\imath(\omega-\omega_0)t}-H.C\right]\nonumber\\
&&+\imath m\omega_0\int_0^\infty d\omega \int_0^\infty d\omega'
\frac{\hbar}{2\sqrt{\omega\omega'}}\
f^{(2)}_{ijk}(\omega,\omega')\left[d_i\sigma b^\dag_j(\omega)
b^\dag_k(\omega')
e^{-\imath(\omega_0-\omega-\omega')t}\right.\nonumber\\
&&+d_i\sigma b_j(\omega) b^\dag_k(\omega')
e^{-\imath(\omega_0+\omega-\omega')t}+d_i\sigma b^\dag_j(\omega)
b_k(\omega') e^{-\imath(\omega_0-\omega+\omega')t}\nonumber\\
&&-d^*_i\sigma^\dag b_j(\omega) b_k(\omega')
e^{\imath(\omega_0-\omega-\omega')t}-d^*_i\sigma^\dag b_j(\omega)
b_k^\dag(\omega') e^{\imath(\omega_0-\omega+\omega')t}\nonumber\\
&&\left. - d^*_i\sigma^\dag b^\dag_j(\omega) b_k(\omega')
e^{\imath(\omega_0+\omega-\omega')t}\right]
\end{eqnarray}
where the symmetry relation (\ref{b11}) has been used. Let  the
combined density operator of the atom together with the environment
is denoted by $\rho_{SR}$ in the interaction picture. Then, the
reduced density operator of the atom alone, denoted by $\rho_S$, is
obtained by taking the trace of $\rho_{SR}$ with respect to the
coordinates of the environment, that is $\rho_S=Tr_R[\rho_{SR}]$.
Since it is assumed that $H'_I(t)$ is sufficiently small, according
to the density operator approach for the damping theory \cite{31},
the time evolution of the reduced density operator $\rho_S$ is the
solution of equation
\begin{eqnarray} \label{c4}
&&\dot{\rho}_s(t)=-\frac{\imath}{\hbar}Tr_R[H'_I(t)\ ,\
\rho_s(0)\otimes\rho_R(0)]\nonumber\\
&&-\frac{1}{\hbar^2}Tr_R\int_0^t dt'\left[H'_I(t),\ ,\
\left[H'_I(t')\ ,\ \rho_s(t)\otimes\rho_R(0)\right]\right]
\end{eqnarray}
 up to order of $H'^2_I$, where $\rho_R(0)$ is the density operator of
 the environment at $t=0$. In this formalism the environment is taken in equilibrium. Also the Markovian approximation has been applied
 replacing
$\rho_S(t')$ by $\rho_S(t)$ in the integrand in Eq.(\ref{c4}).\\
To calculate the spontaneous emission of the atom, the initial
states of the atom and the environment are taken as
\begin{equation}\label{c5}
 \rho_R(0)=|0\rangle\langle0| \hspace{2.00
cm}\rho_S(0)=|2\rangle\langle 2|
\end{equation}
where $|0\rangle$ is the vacuum state of the environment. Now
substituting $H'_I(t)$ from (\ref{c3}) into (\ref{c4}) and regarding
(\ref{c5}) the time evolution of the reduced density operator
$\rho_S$ is obtained as
\begin{eqnarray*}
&&\dot{\rho}_S=\frac{m\omega_0}{2}\int_0^\infty
\frac{d\omega}{\omega}\ f^{(2)}_{ijj}(\omega,\omega)[ d_i\sigma
e^{-\imath\omega_0 t}+H.C]\nonumber\\
&&-\frac{m^2\omega_0^2}{2\hbar} \int_0^\infty \frac{d\omega}{\omega}
\  d^*_i\ f^{(1)}_{ij}(\omega)\  f^{(1)}_{lj}(\omega)\ d_l \left
[\int_0^t dt'\
e^{-\imath(\omega-\omega_0)(t-t')}\ \sigma^\dag\sigma \rho_S(t)+H.C\right]\nonumber\\
&&+\frac{m^2\omega_0^2}{\hbar} \int_0^\infty \frac{d\omega}{\omega}
\  d^*_i\ f^{(1)}_{ij}(\omega)\  f^{(1)}_{lj}(\omega)\ d_l \ \sigma
\rho_S(t) \sigma^\dag \int_0^t dt'\
\cos(\omega-\omega_0)(t-t')\nonumber\\
&&-\frac{m^2\omega_0^2}{2} \int_0^\infty d\omega_1 \int_0^\infty
d\omega_2 \frac{1}{\omega_1 \omega_2}\  d_{i_1}\
f^{(2)}_{i_1j_1j_1}(\omega_1,\omega_1)\
f^{(2)}_{i_2j_2j_2}(\omega_2,\omega_2)\
d_{i_2}\nonumber\\
&&\times\left[\int_0^t dt'\ \sigma \rho_S(t)\ \sigma e^{-\imath
\omega_0(t+t')}+H.C\right]\nonumber\\
&&+\frac{m^2\omega_0^2}{2} \int_0^\infty d\omega_1 \int_0^\infty
d\omega_2 \frac{1}{\omega_1 \omega_2}\  d^*_{i_1}\
f^{(2)}_{i_1j_1j_1}(\omega_1,\omega_1)\
f^{(2)}_{i_2j_2j_2}(\omega_2,\omega_2)\ d_{i_2} \nonumber\\
&&\times \ [\sigma^\dag \rho_S(t)\ \sigma +\sigma \rho_S(t)\
\sigma^\dag] \int_0^t
dt'\cos\omega_0(t-t')\nonumber\\
&&+m^2\omega_0^2 \int_0^\infty d\omega_1 \int_0^\infty d\omega_2
\frac{1}{\omega_1 \omega_2}\  d^*_{i_1}\
f^{(2)}_{i_1j_1j_2}(\omega_1,\omega_2)\
f^{(2)}_{i_2j_1j_2}(\omega_1,\omega_2)\ d_{i_2}\nonumber\\
&&\times \sigma \rho_S(t)\ \sigma^\dag \int_0^t
dt'\cos(\omega_0-\omega_1-\omega_2)(t-t')\nonumber\\
&&-\frac{m^2\omega_0^2}{4} \int_0^\infty d\omega_1 \int_0^\infty
d\omega_2 \frac{1}{\omega_1 \omega_2}\  d^*_{i_1}\
f^{(2)}_{i_1j_1j_1}(\omega_1,\omega_1)\
f^{(2)}_{i_2j_2j_2}(\omega_2,\omega_2)\ d_{i_2} \nonumber\\
&&\times\left[\int_0^t dt'\
e^{-\imath\omega_0(t-t')}\ \sigma\sigma^\dag \rho_S(t)+H.C\right]\nonumber\\
\end{eqnarray*}
\begin{eqnarray*}
&&-\frac{m^2\omega_0^2}{4} \int_0^\infty d\omega_1 \int_0^\infty
d\omega_2 \frac{1}{\omega_1 \omega_2}\  d^*_{i_1}\
f^{(2)}_{i_1j_1j_1}(\omega_1,\omega_1)\
f^{(2)}_{i_2j_2j_2}(\omega_2,\omega_2)\ d_{i_2} \nonumber\\
&&\times\left[\int_0^t dt'\
e^{-\imath\omega_0(t-t')}\ \rho_S(t)\sigma^\dag\sigma+H.C\right]\nonumber\\
&&-\frac{m^2\omega_0^2}{2} \int_0^\infty d\omega_1 \int_0^\infty
d\omega_2 \frac{1}{\omega_1 \omega_2}\  d^*_{i_1}\
f^{(2)}_{i_1j_1j_2}(\omega_1,\omega_2)\
f^{(2)}_{i_2j_1j_2}(\omega_1,\omega_2)\ d_{i_2} \nonumber\\
&&\times\left[\int_0^t dt'\
e^{-\imath(\omega_0-\omega_1-\omega_2)(t-t')}\ \sigma^\dag\sigma
\rho_S(t)+H.C\right]
\end{eqnarray*}
where the the repeated indices implies that the summation should be
done over them. Then, the equation of motion of the matrix elements
$\rho_{S11}=\langle 1|\rho_S| 1\rangle$ , $ \rho_{S22}=\langle
2|\rho_S| 2\rangle$ and $ \rho_{S12}=\rho^*_{S21}=\langle 1|\rho_S|
2\rangle$  now is obtained as
\begin{eqnarray}\label{c7}
&&\dot{\rho}_{S11}=\frac{m^2\omega_0^2}{\hbar} \int_0^\infty
\frac{d\omega}{\omega} \  d^*_i\ f^{(1)}_{ij}(\omega)\
f^{(1)}_{lj}(\omega)\ d_l\ \rho_{S22}(t)\int_0^t dt'\
\cos(\omega-\omega_0)(t-t')\nonumber\\
&&+m^2\omega_0^2 \int_0^\infty d\omega_1 \int_0^\infty d\omega_2
\frac{1}{\omega_1\omega_2}\  d^*_{i_1}\
f^{(2)}_{i_1j_1j_2}(\omega_1,\omega_2)\
f^{(2)}_{i_2j_1j_2}(\omega_1,\omega_2)\ d_{i_2}\
\nonumber\\
&&\times \rho_{S22}(t)\int_0^t dt'\
\cos(\omega_0-\omega_1-\omega_2)(t-t')\nonumber\\
&&+\frac{m^2\omega_0^2}{2} \int_0^\infty d\omega_1 \int_0^\infty
d\omega_2 \frac{1}{\omega_1\omega_2}\  d^*_{i_1}\
f^{(2)}_{i_1j_1j_1}(\omega_1,\omega_1)\
f^{(2)}_{i_2j_2j_2}(\omega_2,\omega_2)\ d_{i_2}\
\rho_{S22}(t)\nonumber\\
&&\times[\rho_{S22}(t)-\rho_{S11}(t)]\int_0^t dt'\ \cos
\omega_0(t-t')
\end{eqnarray}
\begin{eqnarray}\label{c8}
&&\dot{\rho}_{S22}=-\frac{m^2\omega_0^2}{\hbar} \int_0^\infty
\frac{d\omega}{\omega} \  d^*_i\ f^{(1)}_{ij}(\omega)\
f^{(1)}_{lj}(\omega)\ d_l\ \rho_{S22}(t)\int_0^t dt'\
\cos(\omega-\omega_0)(t-t')\nonumber\\
&&-m^2\omega_0^2 \int_0^\infty d\omega_1 \int_0^\infty d\omega_2
\frac{1}{\omega_1\omega_2}\  d^*_{i_1}\
f^{(2)}_{i_1j_1j_2}(\omega_1,\omega_2)\
f^{(2)}_{i_2j_1j_2}(\omega_1,\omega_2)\ d_{i_2}\
\nonumber\\
&&\times \rho_{S22}(t)\int_0^t dt'\
\cos(\omega_0-\omega_1-\omega_2)(t-t')\nonumber\\
&&-\frac{m^2\omega_0^2}{2} \int_0^\infty d\omega_1 \int_0^\infty
d\omega_2 \frac{1}{\omega_1\omega_2}\  d^*_{i_1}\
f^{(2)}_{i_1j_1j_1}(\omega_1,\omega_1)\
f^{(2)}_{i_2j_2j_2}(\omega_2,\omega_2)\ d_{i_2}\
\rho_{S22}(t)\nonumber\\
&&\times[\rho_{S22}(t)-\rho_{S11}(t)]\int_0^t dt'\ \cos
\omega_0(t-t')
\end{eqnarray}
\begin{eqnarray}\label{c9}
&&\dot{\rho}_{S12}=\dot{\rho}^*_{S21}=\frac{m\omega_0}{2}\int_0^\infty\
\frac{d\omega}{\omega}\ f^{(2)}_{ijj}(\omega,\omega)\ d_i
e^{-\imath\omega_0 t}\nonumber\\
&&-\frac{m^2\omega_0^2}{2}\ \int_0^\infty d\omega_1 \int_0^\infty
d\omega_2 \frac{1}{\omega_1\omega_2}\  d_{i_1}\
f^{(2)}_{i_1j_1j_1}(\omega_1,\omega_1)\
f^{(2)}_{i_2j_2j_2}(\omega_2,\omega_2)\ d_{i_2}\nonumber\\
&&\times \rho_{S22}(t) \int_0^t dt' e^{-\imath\omega_0 (t+t')}
\end{eqnarray}
For sufficiently large times the integrals appeared  in the
equations (\ref{c7}) and (\ref{c8}) can be approximated  by
\begin{eqnarray}\label{c10}
&&\frac{1}{\pi}\int_0^t dt'\ \cos(\omega-\omega_0)(t-t')\sim
\delta(\omega-\omega_0)\nonumber\\
&&\frac{1}{\pi}\int_0^t dt'\
\cos(\omega_0-\omega_1-\omega_2)(t-t')\sim
\delta(\omega_0-\omega_1-\omega_2)\nonumber\\
&&\frac{1}{\pi}\int_0^t dt'\ \cos\omega_0(t-t')\sim
\delta(\omega_0)=0
\end{eqnarray}\label{c11}
Hence the time evolution of the matrix elements $ \rho_{S11}$ and
$\rho_{S22}$ for sufficiently large times is reduced to
\begin{equation}\label{c12}
\dot{\rho}_{S11}=\Gamma \ \rho_{S22}\hspace{2.00
cm}\dot{\rho}_{S22}=-\Gamma \ \rho_{S22}
\end{equation}
where
\begin{eqnarray}\label{c13}
&& \Gamma=\frac{\pi m\omega_0^2}{\hbar} \ d_i^* f^{(1)}_{ij}(\omega)
f^{(1)}_{lj}(\omega) d_l\nonumber\\
&&+\pi m^2 \omega_0^2\int_0^\infty d\omega\
\frac{1}{\omega(\omega-\omega_0)}\ d^*_{i_1}\
f^{(2)}_{i_1j_1j_2}(\omega,\omega-\omega_0)\
f^{(2)}_{i_2j_1j_2}(\omega,\omega-\omega_0)\ d_{i_2}\nonumber\\
&&
\end{eqnarray}
is the decay rate of the spontaneous emission of the initially
excited two level atom up to the first order of nonlinearity. The
first term in (\ref{c13}) is the decay rate in the absence of
nonlinearity effects and the second term is the first contribution
related to nonlinear effects of the environment.  It may be noted
from (\ref{c12}) that $\dot{\rho}_{S11}+\dot{\rho}_{S22}=0$ which
implies the conservation of the probability. An important point is
that the decay rate $\Gamma$ is invariant under the various coupling
tensors which is related to each other by the transformation
(\ref{b14}). This should be so, because the decay rate $\Gamma$ is a
physical observable.
\section{Summary}
A fully canonical quantization of a quantum system moving in an
anisotropic  non-linear absorbing environment was introduced. The
main dissipative system was coupled with the environment through
some coupling tensors of various ranks. The coupling tensors have an
important role in this theory.  Based on a response equation, the
forces against the motion of the main system were resolved into two
parts,  the damping forces and the noise forces. The response
equation of the environment was obtained using  the Heisenberg
equations describing the time evolution of the coordinates of the
system and the environment. Some susceptibility tensors of various
ranks were attributed to the environment. The susceptibility tensors
in the linear and non-linear regimes  were defined in terms of the
coupling tensors of the system and its environment. It was shown
that, by imposing some symmetry conditions on the coupling tensors,
the susceptibility tensors obey the symmetry properties reported in
the literature. A realization of this quantization method is the
quantized electromagnetic field in the presence of a non-linear
absorbing dielectric. Finally the effect of the nonlinearity of the
environment was investigated on the spontaneous decay rate of  a
two-level atom imbedded in  the non-linear environment.

\end{document}